# Direct electronic measurement of Peltier cooling and heating in graphene

I.J. Vera-Marun[1,2], J.J. van den Berg[1], F.K. Dejene[1] & B.J. van Wees[1]

Thermoelectric effects allow the generation of electrical power from waste heat and the electrical control of cooling and heating. Remarkably, these effects are also highly sensitive to the asymmetry in the density of states around the Fermi energy and can therefore be exploited as probes of distortions in the electronic structure at the nanoscale. Here we consider two-dimensional graphene as an excellent nanoscale carbon material for exploring the interaction between electronic and thermal transport phenomena, by presenting a direct and quantitative measurement of the Peltier component to electronic cooling and heating in graphene. Thanks to an architecture including nanoscale thermometers, we detected Peltier component modulation of up to 15 mK for currents of 20 µA at room temperature and observed a full reversal between Peltier cooling and heating for electron and hole regimes. This fundamental thermodynamic property is a complementary tool for the study of nanoscale thermoelectric transport in two-dimensional materials.

[1] Physics of Nanodevices, Zernike Institute for Advanced Materials, University of Groningen, Nijenborgh 4, 9747 AG Groningen, The Netherlands. [2] School of Physics and Astronomy, The University of Manchester, Schuster Building-2.14, Manchester M13 9PL, UK. Correspondence and requests for materials should be addressed to I.J.V.-M. (email: ivan.veramarun@manchester.ac.uk).

Recent advances in thermoelectrics[1,2] have been fuelled by nanoscaled materials[3,4], with carbon-based ones offering prospects of addressing large power density via heat management and exploiting thermoelectric effects[5–7]. A basic description of thermoelectrics usually involves two reciprocal processes: the Seebeck and Peltier effects. The Seebeck effect is the generation of a voltage due to a temperature difference and is quantified by the Seebeck coefficient or thermopower of a material, $S = -\Delta V/\Delta T$, used for temperature sensing in thermocouples. Graphene[8–10] has been shown theoretically[11–13] and experimentally[14–16] to have a large and tunable $S$ up to $\pm 100\,\mu V\,K^{-1}$ at room temperature, due to its unique electronic band structure and electrostatic tunability of the density and polarity of its charge carriers. In contrast, the Peltier effect describes the heating or cooling of a junction between two different materials when an electric charge current is present. It is quantified by the Peltier coefficient $\Pi$, which can be understood as the heat transported by thermally excited charge carriers. The Peltier effect is a reversible thermodynamic phenomenon that depends linearly on the current, so it is fundamentally different from the irreversible Joule heating[17]. More importantly, as both thermoelectric coefficients are related by the second Thomson relation[18] $\Pi = ST$, where $T$ is the reference temperature, it follows that in graphene also the Peltier coefficient $\Pi$ (and its associated cooling or heating action) can be controlled in both magnitude and sign. Until now, one study managed to detect Peltier heat in a graphene–metal junction[19], nevertheless without demonstrating any significant modulation nor reversal of the Peltier effect with carrier density, and involved a complex scanning probe microscopy technique.

This work presents a direct and quantitative electronic measurement of Peltier cooling and heating, in both single layer (SL) and bilayer (BL) graphene, demonstrating full modulation of the Peltier effect via electrostatic gating. We use nanoscale thermocouples for a sensitive and spatially resolved thermometry of the Peltier electronic heat evolved or absorbed at a graphene–metal junction. The results are consistent with the reversibility and electron-hole symmetry expected for the linear response of the Peltier effect. Furthermore, we probe both the local temperature change on the junction where the Peltier effect is induced, as well as in another junction some distance away. We successfully describe the observed temperature profile in the device using a simple one-dimensional model.

## Results

**Device architecture.** We induced the Peltier effect by sending a charge current through a graphene–Au metal junction (Fig. 1). With the current $I$ directed from graphene to Au, the evolution of the Peltier heat at the junction is then given by $\dot{Q} = (\Pi_{gr} - \Pi_{Au})I \approx \Pi_{gr}I$, as for most carrier densities $|\Pi_{gr}| \gg \Pi_{Au}$ (refs 14–16,20). In the hole regime $\Pi_{gr} > 0$, which corresponds to Peltier heating of the junction, as depicted in Fig. 1a. Reversely, a junction where $I$ goes from Au into graphene has a cooling rate of the same magnitude. Finally, in the electron regime $\Pi_{gr} < 0$, reversing the effects of cooling and heating (Fig. 1b).

To probe the electronic temperature of the Peltier junction we used nanoscale NiCu/Au thermocouple junctions ($S_{NiCu} \approx -30\,\mu V\,K^{-1}$ and $S_{Au} \approx 2\,\mu V\,K^{-1}$, see Methods, Supplementary Fig. 1 and Supplementary Note 1), placed outside the graphene channel but in close proximity to the Peltier junction. The thermocouple builds up an open circuit potential $V_{tc} = (S_{NiCu} - S_{Au})\Delta T = S_{tc}\Delta T$ (between contacts 3 and 4 in Fig. 2) when a temperature difference $\Delta T$ exists at the thermocouple junction with respect to the reference temperature $T$. This sensitive nanoscale thermometry can detect temperature changes in the mK regime[21].

Most importantly, this approach does not require any charge current present in the thermocouple detection circuit, making it compatible with the requirement of applying a current through the graphene–metal junction for the generation of the Peltier effect. This is in contrast to the resistive thermometry used in the standard architecture for the Seebeck effect[14–16] where a sensing current along the resistor is needed.

**Themoelectric measurements.** For the electrical generation of the Peltier effect we applied a low frequency AC current $I$ of up to $20\,\mu A$ to the graphene–metal junction (between contacts 1 and 2 of Fig. 2) and used a lock-in technique to measure the thermocouple voltage. With this technique, we can distinguish between Peltier ($\propto I$) and higher-order contributions such as Joule heating ($\propto I^2$) by separating the first harmonic response to the heat modulation at the junction. From the second harmonic, we estimate that Joule heating at $20\,\mu A$ is $\sim 10\,mK$ at $300\,K$, similar to the Peltier cooling and heating. Our measurement scheme allows us to single out the Peltier component, excluding all other possible sources from the measured signal (see Methods, Supplementary Fig. 2 and Supplementary Note 2) and realizing a complementary tool for the study of nanoscale thermoelectric transport in two-dimensional materials. Here we quantify the Peltier signal by normalizing the voltage generated at the nanoscale thermocouple by the current driving the Peltier junction, $V_{tc}/I$. Thus, our measurement scheme consists of a graphene channel circuit that generates a heat current via the Peltier effect and a nonlocal detector circuit that converts this heat current back into a charge voltage via the Seebeck effect[20].

We observed a modulation in the thermocouple signal $V_{tc}/I \approx 10\,m\Omega$ when changing the carrier density in SL graphene with the use of a backgate potential $V_g$ (see Fig. 3). This corresponds to a modulation of the Peltier coefficient $\Pi_{gr}$. First, we consider the measurement configuration shown in Fig. 2a, with the current direction defined from graphene to metal and the thermocouple electrode grounded. For this configuration we observed a clear change in polarity in $V_{tc}/I$, indicating a reversal of the Peltier effect between heating ($V_g < 5\,V$) and cooling ($V_g > 5$ V). This is consistent with the location of the charge neutrality (Dirac) point (see inset in Fig. 4a) and with the symmetric band structure in graphene. In addition, we consider a reversed configuration with connections to the current source exchanged, such that now the current direction is defined from metal to graphene and the electrode grounded is not the one with the measured thermocouple. This leads to a mirroring of the signal around $V_{tc} = 0$, consistent with the reversible nature of the Peltier effect. The resulting temperature modulation $\Delta T = S_{tc} V_{tc}$ due to the Peltier effect at the graphene–metal junction was $\approx 8$ mK (Fig. 3, right axis).

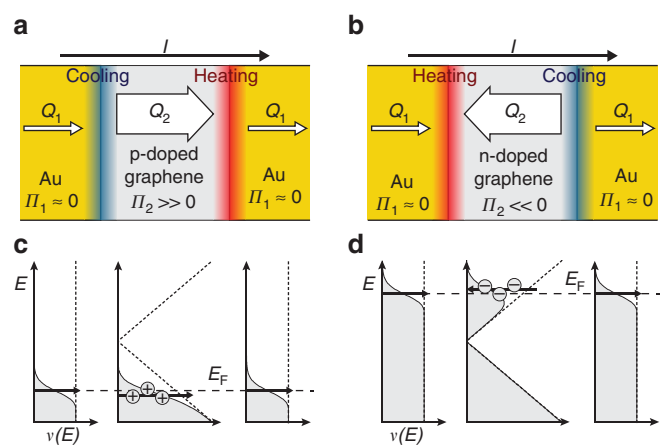

**Figure 1 | Depiction of the Peltier effect at a graphene–Au interface.**
(**a**,**b**) Graphene (grey) has a larger Peltier coefficient $\Pi$ than Au (yellow) and thus a current $I$ can carry more heat in graphene ($Q_2$) than in Au ($Q_1$). As charge flow is conserved, heat is accumulated (red) or absorbed (blue) at the interfaces. (**c**,**d**) The large $\Pi$ in graphene is caused by the strong variation of the density of states $v(E)$ around the Fermi energy $E_F$, lowering or elevating the average energy of the thermalized carriers (indicated by the black arrows). The effect reverses when tuning the carriers from (**a**,**c**) the hole regime to (**b**,**d**) the electron regime.

## Discussion

For a better understanding of the data we calculate $\Pi_{gr}$ from independent charge transport measurements and then use a simple heat balance to describe the temperature modulation $\Delta T$ at the graphene–metal junction. We relate $\Delta T$ to the Peltier heating and cooling rate $\dot{Q}$ via,

$$\Delta T = \dot{Q} R_{th} = \Pi_{gr} I R_{th}, \quad (1)$$

with $R_{th}$ the thermal resistance sensed by the Peltier heat source at the junction, given by the heat transport through the graphene channel and Au electrode, plus the heat flow away to the Si substrate via the $SiO_2$ insulator. As in graphene the thermal conductivity $\kappa_{gr}$ is dominated by phonons[6,7], $R_{th}$ is a constant scaling parameter independent of $V_g$. In contrast, $\Pi_{gr}$ dominates the line shape of the response. To calculate $\Pi_{gr}$, we employ the semi-classical Mott relation[22] together with the density of states for SL graphene, $v(E) = 2E/\pi(\hbar v_F)^2$, to obtain the thermopower[14]. Considering the second Thomson relation, this leads to:

$$\Pi_{gr}^{SL} = \frac{\pi^2 k_B^2 T^2}{3|e|} \frac{1}{G} \frac{dG}{dV} \sqrt{\frac{|e|}{C_g \pi}} \frac{2}{\hbar v_F} \sqrt{|V_g - V_D|}, \quad (2)$$

with $k_B$ the Boltzmann constant, $e$ the electron charge, $\hbar$ the reduced Planck's constant, $v_F$ the Fermi velocity, $C_g = \epsilon_0 \epsilon_r / t_{ox}$ the gate capacitance per unit area, with $t_{ox}$ the $SiO_2$ thickness, $\epsilon_0$ and $\epsilon$ the free-space and relative permittivities, respectively, and $G$ the measured charge conductivity from the Dirac curve.

Figure 4a compares the line shape of $\Delta T$ estimated using equations (1) and (2) with the Peltier measurement from Fig. 3, where we fit the thermal resistance parameter $R_{th}$ with a value to allow a direct comparison at large $V_g$. The good agreement between the two only deviates near the peak in the hole regime. This is because the Peltier effect probes the local density of states at the graphene–metal junction. Therefore, it is much more sensitive to doping from the contact than the Dirac curve (shown in the inset) of the graphene region in between the contacts, which only shows a small electron-hole asymmetry[23]. This observation is consistent with our previous work on nonlinear detection of spin currents in graphene[24], where we have observed a similar modulation in the line shape of a thermoelectric-like response due to contact doping.

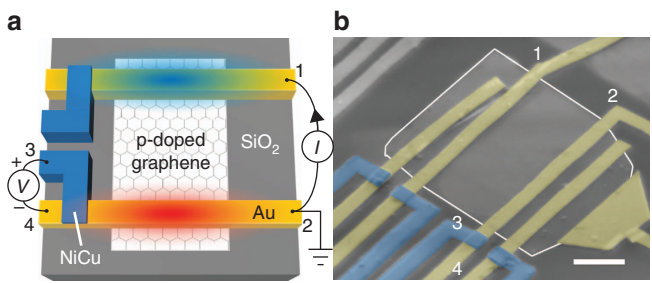

**Figure 2 | Device architecture and measurement configuration.** (**a**) Schematic of the measurement geometry. A graphene flake (white) on a Si/SiO$_2$ substrate is contacted with Au leads (yellow). NiCu leads (blue) form thermocouples to probe the temperature of the graphene–metal interface. We define the current as sent from contact 1 to 2 and probe the thermocouple at contacts 3 and 4. (**b**) Coloured electron micrograph of an actual device. Scale bar, 1 µm. The graphene flake is outlined in white for clarity.

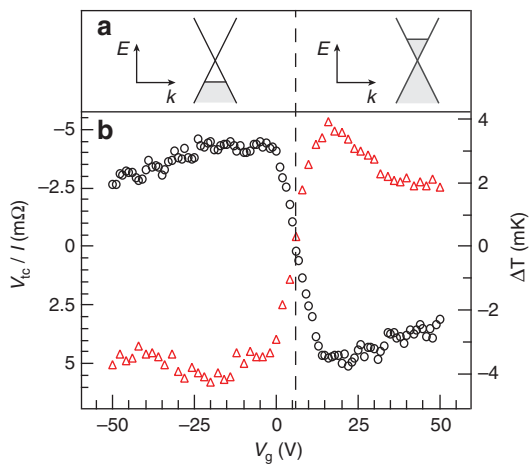

**Figure 3 | Peltier effect in a forward and reverse biased graphene–Au interface.** (**a**) Schematic of the linear dispersion for p-doped (left) and n-doped (right) SL graphene. (**b**) Measurement of the thermocouple signal $V_{tc}/I$ as a function of backgate voltage, for $I = 20$ µA, with the configuration shown in Fig. 1a (black circles). $V_{tc}$ is converted into a temperature by using the Seebeck coefficient of the thermocouple $S_{tc}$, as shown in the right axis. The hole regime shows heating, whereas the electron regime shows cooling. The vertical dashed line indicates the position of the Dirac point. For a reversed configuration of the current source (red triangles) the effects of cooling and heating are reversed.

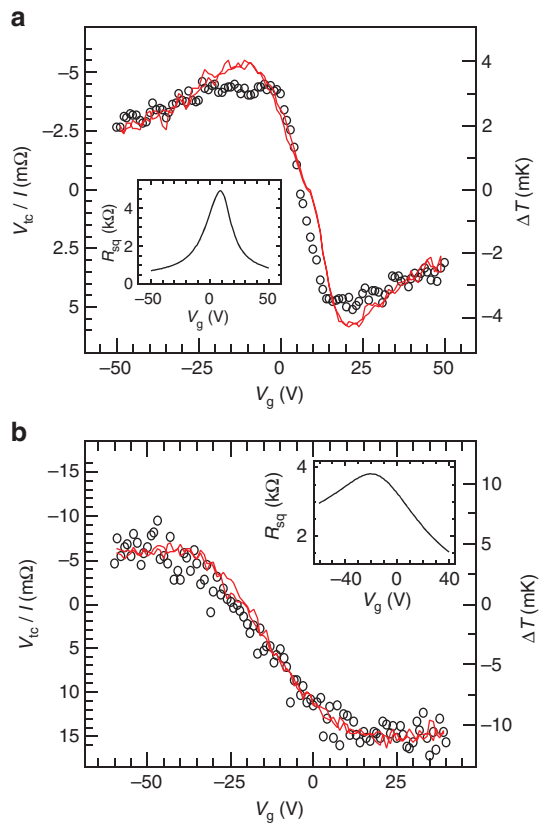

**Figure 4 | Comparison of the Peltier effect in SL and BL graphene.** (**a**) Comparison of Peltier measurement for SL graphene (black circles), with a calculation of the Peltier effect line shape (red lines) derived from charge transport (inset, black lines), using equations (1) and (2). (**b**) Similar comparison for BL graphene, with a calculation using equations (1) and (3).

Figure 4b shows measurements of Peltier cooling and heating in a BL graphene device. We observed the characteristic transition from heating in the hole regime towards cooling in the electron regime, with a temperature modulation of $\sim 15\,\text{mK}$. The transition, located at $V_\text{g} \approx -25\,\text{V}$, correlates with the observed charge neutrality point at $V_\text{D} = -20\,\text{V}$ in the charge transport (see inset). The nonmonotonic behaviour of the Peltier signal is visible for the electron regime, but the parabolic dispersion in BL graphene leads to a broader Peltier curve than for SL graphene. We apply a similar approach as before, to estimate the temperature at the BL graphene–metal junction. Here we use the density of states of BL graphene, $v(E) = 2m/(\pi\hbar^2)$, together with the semiclassical Mott relation, leading to:

$$\Pi^\text{BL}_\text{gr} = \frac{2\pi m k_\text{B}^2 T^2}{3\hbar^2} \frac{1}{G}\frac{dG}{dV}\frac{1}{C_\text{g}}, \qquad (3)$$

with $m \approx 0.05 m_\text{e}$, where $m_\text{e}$ is the free electron mass[10]. The modelled line shape, shown in Fig. 4b, is again scaled by fitting the thermal resistance parameter $R_\text{th}$. We observed an overall agreement between the data and the model, with a lower Peltier signal in the hole regime being consistent with the broader Dirac curve.

A quantitative understanding requires estimating the magnitude of the thermal resistance $R_\text{th}$. Given the geometry of the devices, this usually involves detailed numerical thermal models. To offer physical insight we use a simple one-dimensional model for the heat flow along graphene, with a non-conserved heat current as it flows away via the SiO$_2$ insulator into the Si substrate acting as a thermal reservoir (Fig. 5a). Here we introduce the concept of a thermal transfer length $L_\text{tt}$, defined as the average distance heat flows along the graphene channel (Fig. 5b). It is given by[25] $L_\text{tt} = \sqrt{\kappa_\text{gr} t_\text{gr} t_\text{ox}/\kappa_\text{ox}}$, with $\kappa_\text{gr}$ ($\kappa_\text{ox}$) the thermal conductivity and $t_\text{gr}$ ($t_\text{ox}$) the thickness of graphene (SiO$_2$). Considering the thermal conductivity $\kappa_\text{gr} = 600\,\text{W m}^{-1}\,\text{K}^{-1}$ for SL graphene supported on a Si/SiO$_2$ substrate[26], which is reduced from its intrinsic value due to substrate coupling, we estimate $L_\text{tt} \approx 320\,\text{nm}$. The small value indicates that the temperature modulation due to the Peltier effect diffuses laterally a short distance from the contact. With this characteristic length, we can readily estimate the thermal resistance of a heat transport channel in analogy to the study of spin resistance[24] (see Methods). The estimated $R_\text{th} \approx 1 \times 10^5\,\text{K W}^{-1}$ from the one-dimensional description serves as an upper limit for the thermal resistance, in agreement with the one order of magnitude lower scaling parameter $R_\text{th} \approx 1 \times 10^4\,\text{K W}^{-1}$ used for fitting the modelled curves in Fig. 4.

Finally, we mention two other tests that shed further light on the Peltier origin of the signal. First, Fig. 5c compares the temperature measurement at the Peltier junction in SL graphene with a new measurement where we probed another thermocouple in an adjacent contact, separated by a distance $L = 280\,\text{nm}$. Thus, we can validate the estimated $L_\text{tt}$, as the new measurement should be lower by a factor $e^{-L/L_\text{tt}} \approx 0.4$. The result in Fig. 5b agrees with this estimation. A second test consisted of repeating the measurement from Fig. 3 at 77 K, where we expect the temperature dependence to be dominated by the scaling of the Peltier coefficient, $\Pi_\text{gr} \propto T^2$. The result, a signal one order of magnitude lower (Supplementary Fig. 3 and Supplementary Note 3) further confirms the thermoelectric origin of the response.

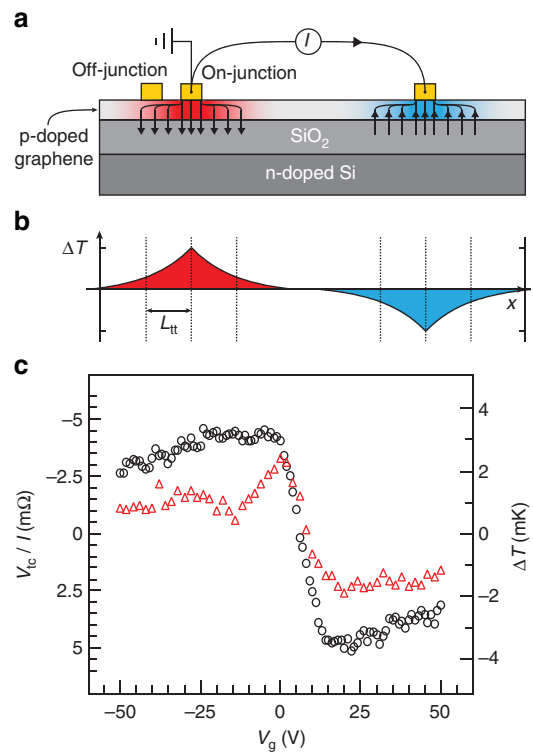

Figure 5 | Temperature profile and off-junction measurement. (a) Schematic cross-section of thermal transport due to Peltier effect (not to scale). (b) Temperature profile in graphene, with characteristic length $L_\text{tt}$ mentioned in the text. (c) Peltier measurement in SL graphene for a thermocouple contacting the graphene–metal junction where the Peltier effect is induced (black circles), and for a thermocouple on an adjacent contact 280 nm away from the junction (red triangles).

Direct measurement of the Peltier effect offers a complementary approach to the study of nanoscale thermoelectric transport in graphene and related two-dimensional materials. Besides providing additional control in electronic heat management at the nanoscale[5–7], quantifying the Peltier effect is useful for studying fundamental thermodynamic relations. In particular, nonlocal measurements involving heat, spin and valley degrees of freedom[24,27–30] have ignored the possibility of a linear Peltier contribution, which will always be present, even without an external magnetic field.

## Methods

**Sample fabrication.** SL and BL graphene flakes were mechanically exfoliated on a Si/SiO$_2$ substrate. To fabricate the device geometry shown in Fig. 2 we used electron beam lithography. First, we deposited using electron beam evaporation Ti (5 nm)/Au (45 nm) electrodes to create ohmic contacts to graphene. The Si substrate was used as a backgate electrode to control the carrier density through a SiO$_2$ dielectric of thickness $t_\text{ox} = 500\,\text{nm}$. Next, after a short cleaning step of the Au surface using Ar ion beam etching, we deposited using sputtering NiCu electrodes to form nanoscale NiCu/Au thermocouples in close proximity to the graphene–metal Peltier junction. We selected NiCu for its large Seebeck coefficient of $S_\text{NiCu} \approx -30\,\mu\text{V K}^{-1}$ (see Supplementary Fig. 1 for an independent measurement)[21] to be used as a thermometer and Au as a contact electrode for the Peltier junction because of its small thermoelectric response ($|\Pi_\text{gr}| \gg \Pi_\text{Au}$) with a Seebeck coefficient of only[20] $S_\text{Au} \approx 2\,\mu\text{V K}^{-1}$. All measured devices (two BLs and one SL) showed consistent results and had typical dimensions of a few micrometres. In this study, we present results for a SL with a channel width of $w^\text{SL}_\text{gr} = 3.0\,\mu\text{m}$ and a BL with $w^\text{BL}_\text{gr} = 3.6\,\mu\text{m}$.

**Peltier measurement.** The measurement of the Peltier effect in a graphene transistor involved the challenge of a sensitive and local thermometry, for which the nanoscale thermocouples were developed. To achieve sub-mK resolution we required the measurement of thermocouple responses in the order of $V_\text{tc}/I \approx 1\,\text{m}\Omega$.

Therefore, we established a careful measurement protocol to differentiate the Peltier response from extrinsic effects.

We applied low amplitude AC currents $I \leq 20\,\mu A$ to the graphene–metal junctions (between Au contacts 1 and 2 of Fig. 2) to keep the response in the linear regime. We used a lock-in technique to separate the first harmonic response to the heat modulation at the graphene–metal junction, to determine the contribution due to Peltier cooling and heating ($\propto I$).

Owing to the finite common mode rejection ratio of the electronics, a local resistance of order $1\,k\Omega$ can lead to a response of the order of $10\,m\Omega$, even for a differential nonlocal measurement. Therefore, all measurements were performed for the sensing configuration shown in Fig. 2, where we measured $V_{34} = V_3 - V_4$, and then repeated for a configuration where the voltage detectors were reversed, $V_{43}$. This allowed us to extract the true differential mode signal $V_{DM} = (V_{34} - V_{43})/2$ and to exclude common mode contributions (see Supplementary Fig. 2).

The frequency $f$ of the AC current was kept low to avoid contributions due to capacitive coupling, for example, between the leads and the backgate. To exclude this contribution, the Peltier lineshape was checked at several frequencies, with a consistent lineshape typically observed for $f \leq 10\,Hz$. All measurements shown here are for $f \leq 3\,Hz$. Small offsets of about $1\,m\Omega$ were corrected by measuring the frequency dependence in the range $0.5-5\,Hz$ and extrapolating to 0 Hz (see Supplementary Fig. 2).

Finally, we directly measured the Seebeck coefficient of the NiCu/Au thermocouples via an independent device geometry (see Supplementary Fig. 1). The result, $S_{tc} = S_{NiCu} - S_{Au} = -27\,\mu V\,K^{-1}$, is consistent with previous estimations[21] and was used to convert the thermocouple signal to a temperature modulation via $\Delta T = S_{tc} V_{tc}$.

**Thermal model.** Here we describe a simplified heat transport model that allows us to estimate the thermoelectric Peltier response and understand its dependence on material parameters analytically. In this model we consider graphene as a one-dimensional diffusive heat transport channel, where the Peltier junctions are treated as point sources for heat currents and the $SiO_2$ substrate acts as a path for heat flow from the graphene into the underlying Si thermal bath. The one-dimensional diffusive description is appropriate, as the $SiO_2$ insulator, with $t_{ox} = 500\,nm$, dominates the thermal transport and is smaller than the width $w_{gr}$ of the graphene channel[25]. Treating the junctions as point sources and disregarding current crowding effects is valid because of the narrowness of the Au contacts, where $w_{Au} \leq 500\,nm \approx L_{ct}$, with $L_{ct}$ being the transfer length for charge transport between the contacts and graphene[19].

Notably, the model is analogous to models commonly used to describe diffusive spin transport[24]. Therefore, it offers physical insight regarding the magnitude of the thermal resistance seen by the Peltier junction, which, in analogy to the study of spin resistance, yields

$$R_{th}^{gr} = \frac{L_{tt}}{2\kappa_{gr} w_{gr} t_{gr}} \quad (4)$$

for heat transport along the graphene channel. Here, $L_{tt} = \sqrt{\kappa_{gr} t_{gr} t_{ox}/\kappa_{ox}}$ is the thermal transfer length introduced in the main text, with $\kappa_{ox} = 1\,W\,m^{-1}\,K^{-1}$ and $\kappa_{gr} = 600\,W\,m^{-1}\,K^{-1}$ (ref. 26). This heat balance only takes into account heat transport along the graphene and the substrate. In practice, there is also transversal heat dissipation through the leads. To account for the latter, we apply the same model above to heat transport across the Au leads and calculate an analogous thermal resistance $R_{th}^{Au}$. For typical device geometries, we obtain $R_{th}^{Au} \approx R_{th}^{gr} \approx 2\times 10^5\,K\,W^{-1}$. We then estimate the total thermal resistance $R_{th} = R_{th}^{gr} \| R_{th}^{Au} \approx 1 \times 10^5\,K\,W^{-1}$, to account for the heat balance of equation (1).

It is noteworthy that the distance between the NiCu/Au thermocouples and the graphene channel ($\approx 500\,nm$) is smaller than the thermal transfer length of the Au leads, $L_{tt}^{Au} = 1.7\,\mu m$, as Au is a good thermal conductor ($\kappa_{Au} = 127\,W\,m^{-1}\,K^{-1}$). Therefore, there is only a correction of 30% to account for the detection efficiency of the thermocouples. We note that this model can only obtain an order of magnitude estimate. It serves as an upper limit for the actual thermal resistance, because it neglects increased lateral heat spreading near the graphene edges and the finite width of the contacts. Finally, using the Wiedemann–Franz law we calculate that electrons only contribute up to 5% to the thermal conductivity of supported graphene, validating our treatment of $R_{th}^{gr}$ as a constant parameter.

## References

1. Bell, L. E. Cooling, heating, generating power, and recovering waste heat with thermoelectric systems. *Science* **321,** 1457–1461 (2008).
2. MacDonald, D. K. C. *Thermoelectricity: An Introduction to the Principles* (Dover Publications, 2006).
3. Cho, S. *et al.* Thermoelectric imaging of structural disorder in epitaxial graphene. *Nat. Mater.* **12,** 913–918 (2013).
4. Heremans, J. P., Dresselhaus, M. S., Bell, L. E. & Morelli, D. T. When thermoelectrics reached the nanoscale. *Nat. Nanotechnol.* **8,** 471–473 (2013).
5. Dubi, Y. & Di Ventra, M. Colloquium: heat flow and thermoelectricity in atomic and molecular junctions. *Rev. Mod. Phys.* **83,** 131–155 (2011).
6. Balandin, A. A. Thermal properties of graphene and nanostructured carbon materials. *Nat. Mater.* **10,** 569–581 (2011).
7. Pop, E., Varshney, V. & Roy, A. K. Thermal properties of graphene: fundamentals and applications. *MRS Bull.* **37,** 1273–1281 (2012).
8. Novoselov, K. S. *et al.* Electric field effect in atomically thin carbon films. *Science* **306,** 666–669 (2004).
9. Castro Neto, A. H., Guinea, F., Peres, N. M. R., Novoselov, K. S. & Geim, A. K. The electronic properties of graphene. *Rev. Mod. Phys.* **81,** 109–162 (2009).
10. Das Sarma, S., Adam, S., Hwang, E. H. & Rossi, E. Electronic transport in two-dimensional graphene. *Rev. Mod. Phys.* **83,** 407–470 (2011).
11. Ouyang, Y. & Guo, J. A theoretical study on thermoelectric properties of graphene nanoribbons. *Appl. Phys. Lett.* **94,** 263107 (2009).
12. Yan, X.-Z., Romiah, Y. & Ting, C. S. Thermoelectric power of Dirac fermions in graphene. *Phys. Rev. B* **80,** 165423 (2009).
13. Hwang, E. H., Rossi, E. & Das Sarma, S. Theory of thermopower in two-dimensional graphene. *Phys. Rev. B* **80,** 235415 (2009).
14. Zuev, Y. M., Chang, W. & Kim, P. Thermoelectric and magnetothermoelectric transport measurements of graphene. *Phys. Rev. Lett.* **102,** 096807 (2009).
15. Wei, P., Bao, W., Pu, Y., Lau, C. N. & Shi, J. Anomalous thermoelectric transport of dirac particles in graphene. *Phys. Rev. Lett.* **102,** 166808 (2009).
16. Checkelsky, J. G. & Ong, N. P. Thermopower and Nernst effect in graphene in a magnetic field. *Phys. Rev. B* **80,** 081413 (2009).
17. Giazotto, F., Heikkilä, T. T., Luukanen, A., Savin, A. M. & Pekola, J. P. Opportunities for mesoscopics in thermometry and refrigeration: physics and applications. *Rev. Mod. Phys.* **78,** 217–274 (2006).
18. Callen, H. B. The application of Onsager's reciprocal relations to thermoelectric, thermomagnetic, and galvanomagnetic effects. *Phys. Rev.* **73,** 1349–1358 (1948).
19. Grosse, K. L., Bae, M.-H., Lian, F., Pop, E. & King, W. P. Nanoscale Joule heating, Peltier cooling and current crowding at graphene-metal contacts. *Nat. Nanotechnol.* **6,** 287–290 (2011).
20. Bakker, F. L., Slachter, A., Adam, J.-P. & van Wees, B. J. Interplay of Peltier and Seebeck effects in nanoscale nonlocal spin valves. *Phys. Rev. Lett.* **105,** 136601 (2010).
21. Bakker, F. L., Flipse, J. & van Wees, B. J. Nanoscale temperature sensing using the Seebeck effect. *J. Appl. Phys.* **111,** 084306 (2012).
22. Cutler, M. & Mott, N. F. Observation of anderson localization in an electron gas. *Phys. Rev.* **181,** 1336–1340 (1969).
23. Nouchi, R. & Tanigaki, K. Charge-density depinning at metal contacts of graphene field-effect transistors. *Appl. Phys. Lett.* **96,** 253503 (2010).
24. Vera-Marun, I. J., Ranjan, V. & van Wees, B. J. Nonlinear detection of spin currents in graphene with non-magnetic electrodes. *Nat. Phys.* **8,** 313–316 (2012).
25. Bae, M.-H., Ong, Z.-Y., Estrada, D. & Pop, E. Imaging, simulation, and electrostatic control of power dissipation in graphene devices. *Nano Lett.* **10,** 4787–4793 (2010).
26. Seol, J. H. *et al.* Two-dimensional phonon transport in supported graphene. *Science* **328,** 213–216 (2010).
27. Abanin, D. A. *et al.* Giant nonlocality near the Dirac point in graphene. *Science* **332,** 328–330 (2011).
28. Renard, J., Studer, M. & Folk, J. A. Origins of nonlocality near the neutrality point in graphene. *Phys. Rev. Lett.* **112,** 116601 (2014).
29. Gorbachev, R. V. *et al.* Detecting topological currents in graphene superlattices. *Science* **346,** 448–451 (2014).
30. Sierra, J. F., Neumann, I., Costache, M. V. & Valenzuela, S. O. Hot-carrier seebeck effect: diffusion and remote detection of hot carriers in graphene. *Nano Lett.* **15,** 4000–4005 (2015).

### Acknowledgements

We thank B. Wolfs, J.G. Holstein, H.M. de Roosz, H. Adema and T. Schouten for their technical assistance. We acknowledge the financial support of the Netherlands Organisation for Scientific Research (NWO), the Zernike Institute for Advanced Materials, the Dutch Foundation for Fundamental Research on Matter (FOM) and the European Union Seventh Framework Programmes ConceptGraphene (No. 257829), Graphene Flagship (No. 604391) and the Future and Emerging Technologies (FET) programme under FET-Open grant number 618083 (CNTQC).

### Author contributions

I.J.V.-M. and B.J.v.W. conceived the experiments. I.J.V.-M., J.J.v.dB. and F.K.D. carried out the sample fabrication, measurements and data analysis. I.J.V.M. and J.J.v.dB. wrote the manuscript. All authors discussed the results and the manuscript.

### Additional information

**Supplementary Information** accompanies this paper at http://www.nature.com/naturecommunications

**Competing financial interests:** The authors declare no competing financial interests.

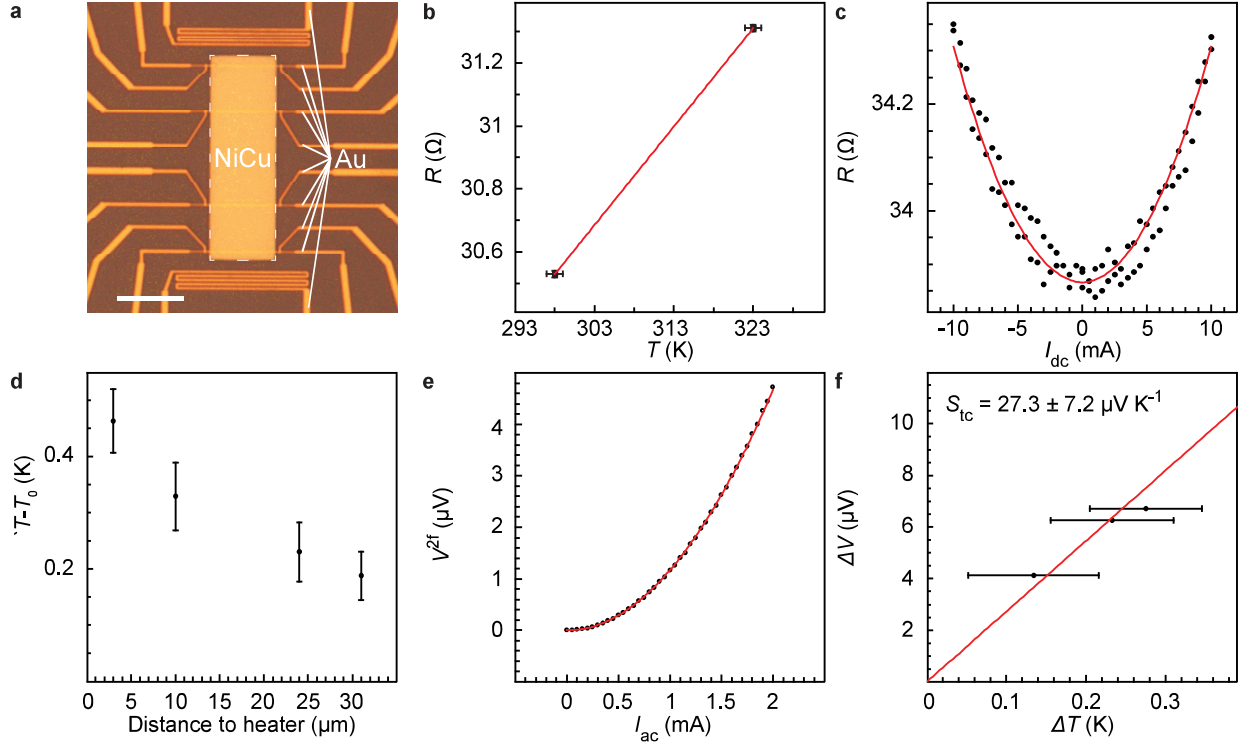

**Supplementary Figure 1. Thermocouple calibration.** (a) Micrograph of the geometry for measuring the Seebeck coefficient $S_{\text{tc}} = S_{\text{NiCu}} - S_{\text{Au}}$. The device consists of a NiCu channel with Au bar electrodes. The microfabricated Au heater creates a temperature gradient which we mapped using the 4-probe resistance of the Au bars. The Au bars were also used as contacts to measure the Seebeck voltage of the NiCu/Au thermocouple. Scale bar is 10 µm. (b) Resistance of a Au bar versus global temperature (black squares). Error bars are the s.d. in the global temperature measurement. For this bar, the linear fit (red line) gives a thermistor sensitivity of $31 \pm 3$ mΩ K$^{-1}$. (c) Dependence of the resistance of a Au bar to a dc current $I_{\text{dc}}$ applied through the microfabricated Au heater (black circles). The quadratic fit (red line) gives the thermistor response in units of Ω A$^{-2}$. (d) Temperature profile along the NiCu film extracted at each Au bar, for a fixed $I_{\text{dc}} = 2$ mA. Error bars are the s.d. in the temperature at each bar propagated from the analysis of fits as those in (b) and (c). (e) Seebeck voltage along the NiCu channel due to an ac heater current (black circles), showing a quadratic ($\propto I^2$) dependence and no voltage offset at $I_{\text{ac}} = 0$ mA (red line). (f) Seebeck voltage $\Delta V$ against the temperature $\Delta T$ (black circles) between pairs of Au bars shown in (d). The linear fit (red line) gives a Seebeck coefficient of $S_{\text{tc}} = -27.3 \pm 7.2$ µV K$^{-1}$.

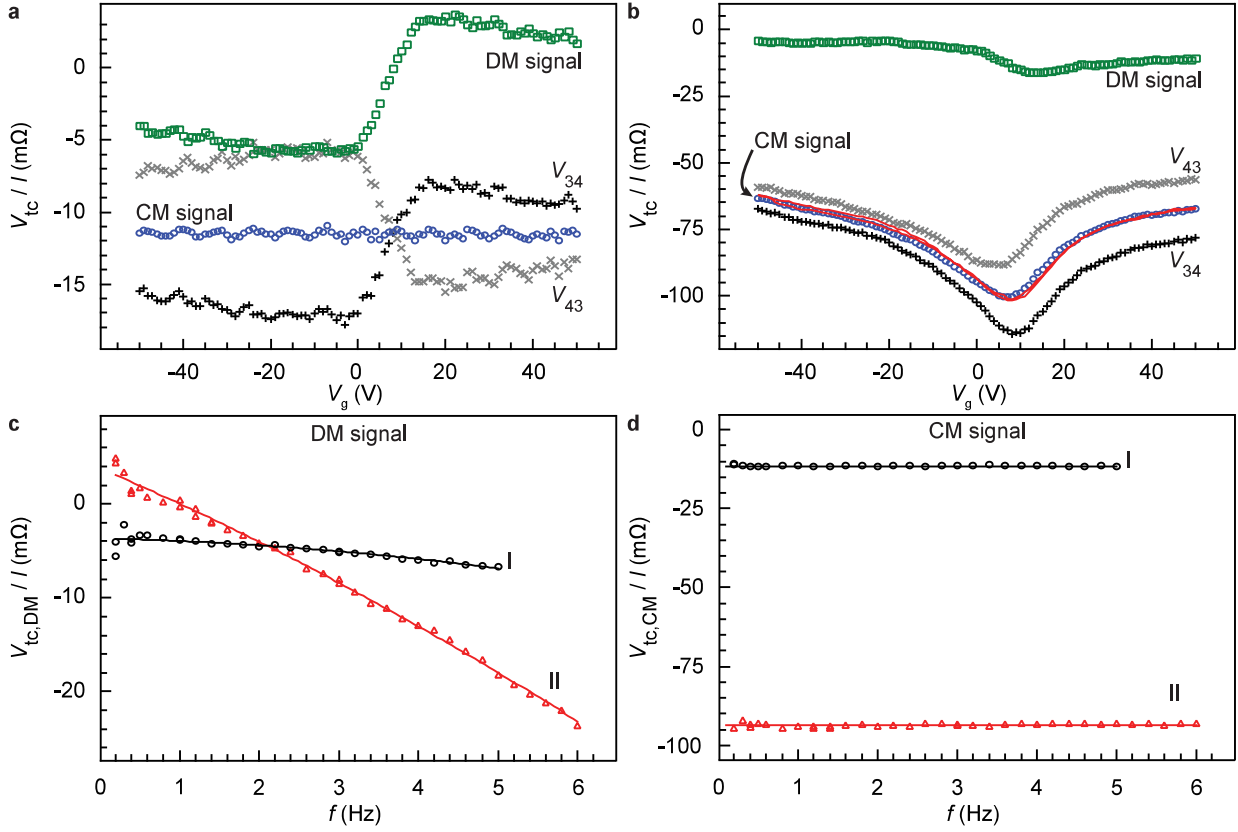

**Supplementary Figure** 2. **Analysis of the thermocouple signal.** (a) Gate dependence of the thermocouple response when a current $I = 20$ µA is applied to the Peltier junction, in the standard current configuration (I) as shown in Fig. 2. The figure shows the raw data, with $V_{34}$ defined as having the voltage probes plus and minus as in Fig. 2 of the main text (black pluses) and $V_{43}$ for the voltage probes reversed (gray crosses). The blue circles show the common mode (CM) and the green squares show the differential mode (DM). This measurement was performed at $f = 3$ Hz. (b) Similar measurements for the reversed current source configuration (II) with different grounding, as described in the main text. In this case, the CM signal is gate dependent due to the contribution of the graphene channel. For a consistency check, we also obtained the source of the CM signal by measuring the resistance $R_3$ as described in the Supplementary Note 1 and multiplying this with the CMMR of the electronics (red line). This measurement was performed at $f = 1$ Hz. (c) Frequency response of the DM mode at $V_g = 0$ in configuration I (black circles) and II (red triangles). The solid lines show the fitting (second order polynomial) functions used to extrapolate to 0 Hz. (d) DM mode for both configurations at $V_g = 0$, which is frequency independent and given by CMRR $\times R_3$ with $R_3$ equal to 1.3 k$\Omega$ (configuration I, black circles) and 10.5 k$\Omega$ (configuration II, red triangles).

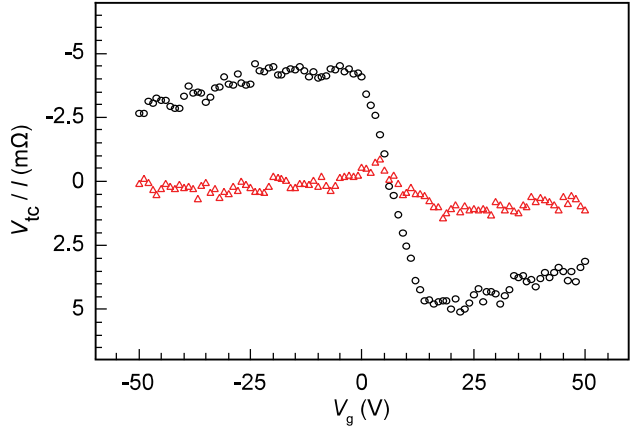

**Supplementary Figure** 3. **Peltier effect at low temperature.** (a) Comparison of Peltier heating and cooling at room temperature (black cirles, as in Fig. 3 of the main text) and at 77 K (red triangles).

## SUPPLEMENTARY NOTES

### Supplementary Note 1. Calibration of the thermocouples

The measurement of a thermocouple voltage can only be related to a local temperature if the Seebeck coefficient $S_{tc}$ of the thermocouple is known. Here, we calibrate our NiCu-Au thermocouples by determining $S_{tc} = S_{NiCu} - S_{Au}$ in a separate device geometry.

Supplementary Figure 1 shows the set of measurements that we used to determine $S_{tc}$ of a thin NiCu film contacted by Au electrodes. For this, we used a microfabricated Au heater and Au resistive thermometers similarly as described in reference[1]. In our case, we sputtered 30 nm of NiCu in a rectangular shape of 31 µm × 10 µm, as shown in Supplementary Figure 1(a). We calibrated the resistive thermometers by measuring the linear temperature dependence of the 4-probe resistance $R_{Au}$ of the Au bars as shown in Supplementary Figure 1(b), using a global heater in the sample mount. Next, we measured the local resistance at each of the Au bars as a function of a dc current $I_{dc}$ through the microfabricated heater. The dependence is quadratic, because in this temperature range $\Delta R_{Au} \propto \Delta T \propto P \propto I_{dc}^2$, where $P$ is the Joule power generated at the heater. A typical measurement with parabolic fit is depicted in Supplementary Figure 1(c). Thus, using the two results above we were able to measure the temperature profile along the NiCu film, as shown in Supplementary Figure 1(d). Finally, measuring the second harmonic response between two different Au bars to an ac current $I_{ac}$ through the microfabricated heater, we show in Supplementary Figure 1(e) the Seebeck potential difference $\Delta V$ induced by the Joule heating. By comparing $\Delta V$ with the temperature difference $\Delta T$ obtained by the previous resistive thermometry, we extract the Seebeck coefficient $S_{\text{NiCu}} - S_{\text{Au}} = -27.3 \pm 7.2$ µV K$^{-1}$ [Supplementary Figure 1(f)]. We note that this value is in good agreement with values reported earlier[2].

**Supplementary Note 2. Analysis of the thermocouple signal**

In order to determine the response caused purely by the Peltier effect it is necessary to carefully analyze the thermocouple response to the current applied to the Peltier junction. In Supplementary Figure 2(a) we show the raw data leading to the signal shown in Fig. 3 of the main text. Here we use the same current source configuration (I) as shown in Fig. 2 of the main text. We define the signal $V_{34}$ as the plus and minus of the voltage probe at contact 3 and 4 respectively, and $V_{43}$ as its reverse. The common mode (CM) signal, given by $(V_{34} + V_{43})/2$, is the background that is caused by any potential relative to ground that is present at both voltage probes, which in this case is constant versus $V_g$ as it only involves the metallic leads. The true differential mode signal (DM) is then given by $(V_{34} - V_{43})/2$. The plot shows that the DM signal is, minus an offset, the true Peltier signal shown in Fig. 3 of the main text.

Supplementary Figure 2(b) shows similar measurements, for the second configuration (II) described in the main text, with exchanged current source connections. Here, the CM signal is gate dependent because there is a graphene resistance between the Peltier junction and the ground. Its origin can be measured independently by applying a current from contact 2 to 1 and measuring the voltage $V_{41}$ (with contact 1 grounded). The red line in Supplementary Figure 2(b) shows the gate dependence of this 3-probe resistance labeled $R_3$, multiplied by the common mode rejection ratio CMRR $= 8.9 \times 10^{-6}$ of the electronics, in good agreement with the CM signal.

We performed all our measurements at low frequency $f < 10$ Hz, ensuring a consistent and frequency independent line shape. The remaining (gate independent) background offset can be evaluated by its frequency dependence. Such background signal comes from capacitive coupling. Therefore, the true value of the Peltier signal is found in the 0 Hz limit. In Supplementary Figure 2(c) is a plot of the $f$ dependence of the DM signals for both measure-ment configurations at $V_\text{g} = 0$ V. Using a second order fit to extrapolate to $f = 0$ Hz we thus obtain the background to substract from the raw DM signals of Supplementary Figure 2(a) and Supplementary Figure 2(b), leading to the final signals of Fig. 3 of the main text. Thus, we confirmed the zero-crossing of the Peltier signal, and thereby the cross-over between heating and cooling, at the charge neutrality point. Supplementary Figure 2(d) shows that the CM signal is $f$ independent for both measurement configurations. Note that, as expected, the smaller contributions of CM and capacitive signals are found for configuration (I), so we focus on this configuration for an accurate analysis.

**Supplementary Note 3. Low temperature measurement**

Thermoelectric coefficients usually have a strong temperature dependence, given they are thermodynamic properties. Therefore, we have repeated the Peltier measurements at liquid nitrogen (LN) temperature. Supplementary Figure 3 compares the measurement as in Fig. 3 of the main text with the same measurement at LN (77 K). At LN we observed a response with a lineshape consistent with that at room temperature (RT), going to more negative values when crossing from the hole to the electron regime, except for an offset of about $-0.5$ m$\Omega$ which lies at the limit of our measurement accuracy. More importantly, the modulation of the signal is about 1 m$\Omega$ or one order of magnitude smaller than at RT.

We can understand this strong suppresion of the signal from Eq. 1 of the main text, which leads to $V_\text{tc}/I = S_\text{tc}\Pi_\text{gr}R_\text{th}$. Considering a linear temperature dependence for the Seebeck coefficients[1], $S \propto T$, and the second Thomson relation[3], $\Pi = ST$, we deduce a scaling of the form $S_\text{tc}\Pi_\text{gr}R_\text{th} \propto T^3 R_\text{th}$. The first factor, $T^3$, leads to a scaling of $(293\text{ K}/77\text{ K})^3 = 55$. An accurate evaluation of $R_\text{th}$ requires detailed modelling and inclusion of several material parameters, but as a first order approximation we consider $R_\text{th} \propto 1/\kappa_\text{gr}$. Since $\kappa_\text{gr}$ has a fivefold decrease from room temperature to LN[4] our final estimate for the scaling, $V_\text{tc}^{RT}/V_\text{tc}^{LN} = 55/5 = 11$, is in agreement with the experimental result, further demonstrating the Peltier thermoelectric nature of the measurements.

## SUPPLEMENTARY REFERENCES

[1] Zuev, Y. M., Chang, W. & Kim, P. Thermoelectric and Magnetothermoelectric Transport Measurements of Graphene. *Phys. Rev. Lett.* **102**, 096807 (2009).

[2] Bakker, F. L., Flipse, J. & Wees, B. J. v. Nanoscale temperature sensing using the Seebeck effect. *J. Appl. Phys.* **111**, 084306 (2012).

[3] Callen, H. B. The Application of Onsager's Reciprocal Relations to Thermoelectric, Thermomagnetic, and Galvanomagnetic Effects. *Phys. Rev.* **73**, 1349–1358 (1948).

[4] Seol, J. H. *et al.* Two-Dimensional Phonon Transport in Supported Graphene. *Science* **328**, 213–216 (2010).